\documentclass{article}
\usepackage{times}
\usepackage{graphicx}
\usepackage{biophysics}

\title{Similarity search for local protein structures at atomic resolution 
by exploiting a database management system}
\author{Akira R. Kinjo\thanks{Corresponding author (akinjo@protein.osaka-u.ac.jp)}~~ and Haruki Nakamura\\
Institute for Protein Research, Osaka University, \\
Suita, Osaka, 565-0871, Japan}
\begin{document}
\maketitle

\begin{abstract}
A method to search for local structural similarities in proteins at atomic 
resolution is presented. 
It is demonstrated that a huge amount of structural data can be handled 
within a reasonable CPU time by using a conventional relational 
database management system with appropriate indexing of geometric data.
This method, which we call geometric indexing, can enumerate 
ligand binding sites that are structurally similar to sub-structures of 
a query protein among more than 160,000 possible candidates within 
a few hours of CPU time on an ordinary desktop computer.
After detecting a set of high scoring ligand binding sites by the geometric
indexing search, structural alignments at atomic resolution are constructed 
by iteratively applying the Hungarian algorithm, and the statistical 
significance of the final score is estimated from an empirical model based 
on a gamma distribution.
Applications of this method to several protein structures clearly 
shows that significant similarities can be detected between local 
structures of non-homologous as well as homologous proteins.
\end{abstract}

\newpage

\section{Introduction}
\label{sec:intro}
According to the `sequence determines structure determines function' paradigm, 
it should be possible to predict protein structure from its amino acid sequence,
and in turn, to predict its function from the structure.
It has been empirically proved, however, that \emph{ab initio} approaches 
to the both of these problems are extremely difficult.
Currently, the most practical and reliable methods for protein structure 
prediction are the ones based on sequence comparison.
In such homology-based methods, sequence similarities imply structural 
similarities.
It is tempting to assume that the same argument applies to the prediction
of protein functions. That is, we expect that we can infer some functional 
information if there are some similarities between two protein structures. 
However, it has been demonstrated that the protein folds (approximate 
over-all structures) of proteins are not significantly correlated with 
their functions.
Since many protein functions such as enzymatic catalysis and ligand binding are 
performed by a small subset of protein atoms or residues, it seems 
necessary to perform local structure comparison in addition to 
(or, instead of) fold comparison for inferring protein function by similarity.

A number of methods have been proposed for searching for local similarities in 
protein structures\cite{JonesANDThornton2004}. However, some of them 
limit the data size due to a prohibitive amount of CPU time and/or RAM 
space required\cite{KinoshitaETAL1999,KinoshitaANDNakamura2003,BrakouliasANDJackson2004}, while others 
sacrifice structural details or diversity for the efficiency of search\cite{WallaceETAL1997,StarkETAL2003,JambonETAL2003}.
The ever increasing structural data in the Protein Data Bank 
(PDB)\cite{wwPDB} include many proteins of unknown functions and hence 
making available efficient and thorough methods for local structure 
comparison for inferring protein functions is a pressing matter. 
At the same time, however, such rapidly increasing data only make conventional 
methods more and more inefficient. It is required that methods for local 
structure comparison be able to follow the rapid increase of data with a 
reasonable scalability.

In this Note, we introduce techniques to construct a scalable method 
for similarity search for local protein structures. 
In this method, ligand binding sites consisting of protein atoms are 
first compiled as a table in a relational database management system 
(RDBMS)\cite{DB_complete}. 
For a given protein structure as a query, the method searches for 
structurally equivalent atoms in the database that 
match the atoms in the query structure. This search process can be 
executed efficiently owing to the indexing mechanism of the RDBMS.
We call this technique \emph{geometric indexing} (GI).
After identifying matching ligand binding sites, alignments at atomic 
resolution are obtained by using the Hungarian algorithm\cite{Lawler,GuptaANDYing1999}.
The present method is similar to the geometric hashing (GH) algorithm in 
spirit. However, since the total size of the structural data 
may well exceed several gigabytes, it is usually not possible to naively 
implement the GH method which must keep a huge hash table in RAM.
On the other hand, an RDBMS stores all the data on a hard disk which is 
much cheaper and larger than RAM, and hence let us overcome the data size 
problem. In addition, almost any modern RDBMS provides an efficient 
indexing mechanism which allows us to retrieve data satisfying a given 
set of constraints rather quickly.
By using the technique introduced here, it becomes possible to keep 
up with the rapidly increasing structural data without sacrificing 
the efficiency of searching or the details and diversity of 
structural information.

\section{Materials and Method}
\label{sec:method}

\subsection{Overview}
\begin{figure}[tb]
  \centering
  \includegraphics[width=8cm]{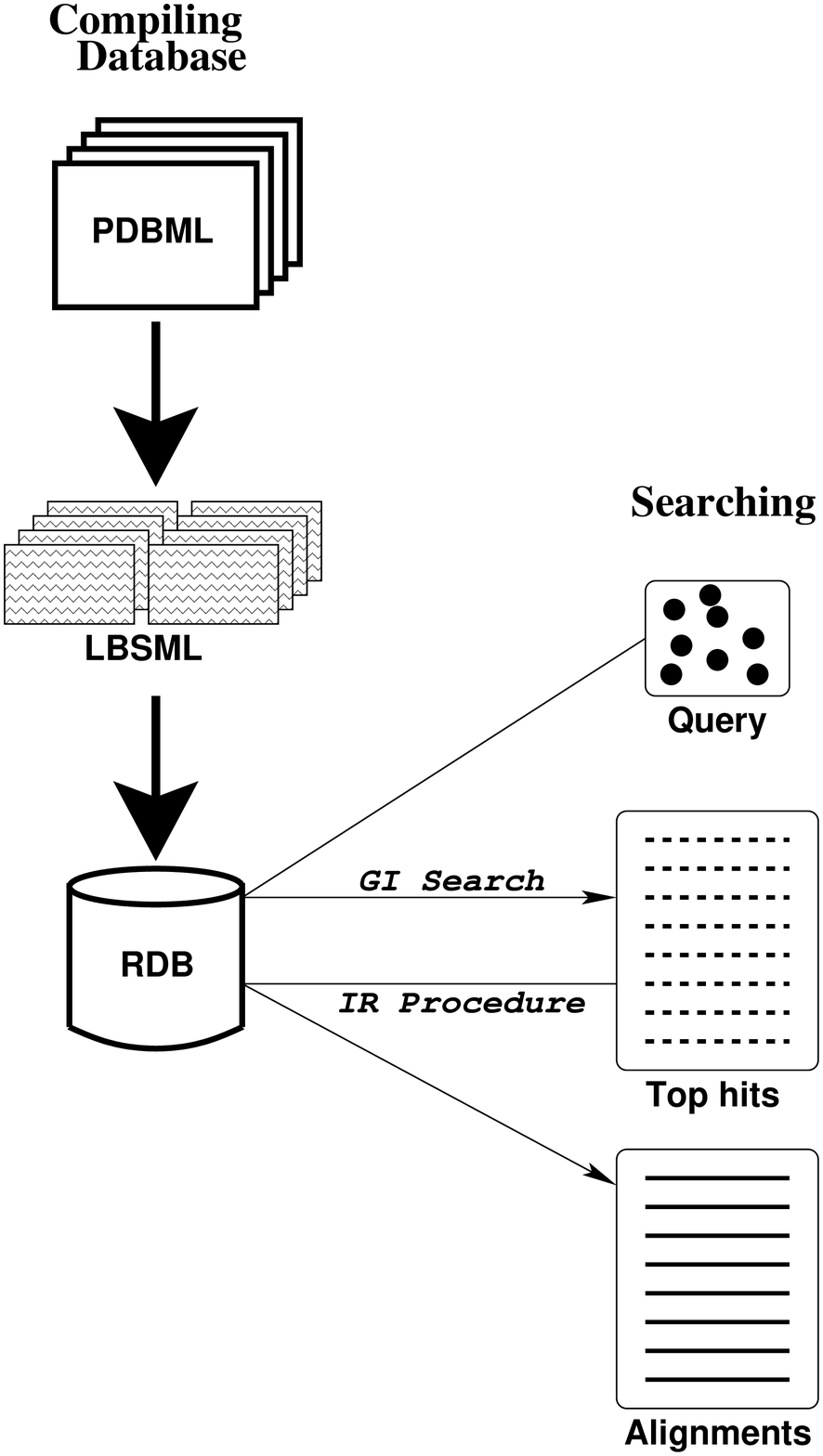}
  \caption{Overview of the method. The left part (``Compiling database'') illustrates the pre-processing step. The right part (``Searching'') shows the search step for a given protein structure as a query.}
  \label{fig:flow}
\end{figure}

We first extract ligand binding sites (templates) from PDBML files\cite{PDBML}
and save them in XML files called LBSML (Ligand Binding Site Markup Language) 
files.
An LBSML file contains information of atoms that are in contact with
a ligand, along with reference sets (refsets) for local coordinate 
systems (see below).
Then we compile refsets and atomic coordinates in local coordinate systems 
into a set of relational database (RDB) tables. 
This is a pre-processing stage and is carried out only once 
as long as we do not need to update the database
(Figure \ref{fig:flow}, left part). 

Then a database search is carried out for a given protein structure as a query
(Figure \ref{fig:flow}, right part). A search is 
divided into two stages. In the first stage, called geometric indexing search 
(``GI Search'' in Figure \ref{fig:flow}), the database is scanned by 
exploiting the indexing mechanism of the RDBMS, and possible atomic 
correspondences are counted.
In the second part (``IR Procedure'' in Figure \ref{fig:flow}), 
a predefined number of high-scoring templates are subject to iterative 
refinement of the alignment to the sub-structures of the query. 

\subsection{Data set}
\label{sec:data}
We downloaded all the PDBML\cite{PDBML} files (43,755 entries) on June 6, 2007.
From these PDB entries, those were discarded that do not contain a protein 
chain or that do not contain any hetero atoms other than water.
\subsection{Definition of reference set (refset)}
\label{sec:refframe}
As in the geometric hashing algorithm, all atomic coordinates are expressed 
in various local coordinate systems defined by reference sets (refsets).
To define refsets, we applied the Delaunay tessellation 
using the Qhull library\cite{Qhull} to each PDB entry. This procedure 
yields a set of tetrahedra consisting of four atoms as the vertices 
that are closest to each other.
Then we selected those tetrahedra whose volumes are between 2 and 10 \AA{}$^3$ 
and whose total accessible areas are greater than zero \AA{}$^2$.
These tetrahedra serve as refsets.
Although only three atoms are necessary to define a unique Cartesian coordinate
system, we use four atoms of a tetrahedron to reduce the number of 
possible combinations for refsets in a later stage of similarity search. 

We define atom types as follows. All the backbone atoms are treated 
uniquely so that backbone ``N'', ``C$_\alpha$'', ``C'' and ``O'' are labeled 
as such and their types are denoted ``BN'', ``BA'', ``BC'', and ``BO'', 
respectively.
The types of side chain atoms are assigned as the corresponding standard 
atom names (as annotated by the ``type\_symbol'' tag of the PDBML file).
We keep only those tetrahedra whose four vertices are of different atom types.
Accordingly, we can lexicographically order the vertices of a tetrahedron 
unambiguously. We can also define the chirality of a tetrahedron (see below). 
Thus, the sequence of ordered atom types and chirality of a tetrahedron
define the type of the tetrahedron. For example, a tetrahedron
consisting of atoms of types ``BN'', ``BA'', ``BC'' and ``S'' with 
positive chirality is typed as ``BA:BC:BC:S:+''.

\begin{figure}[tb]
  \centering
  \includegraphics[width=8cm]{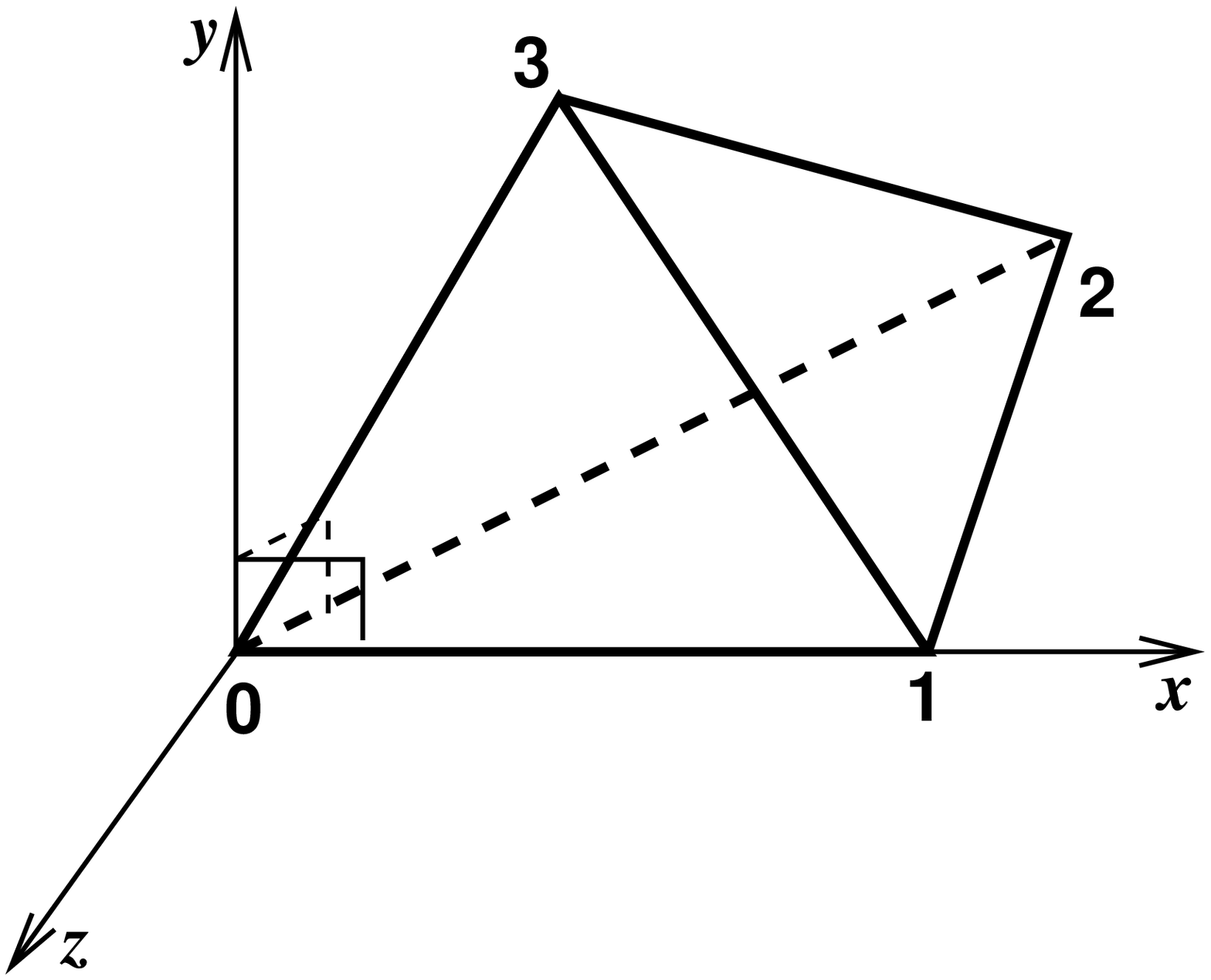}
  \caption{Local coordinate system defined by a refset (tetrahedron).}
  \label{fig:refset}
\end{figure}

Let $\mathbf{r}_{i}$ ($i = 0, \cdots, 3$) be the coordinates of the four atoms 
of a refset (tetrahedron) in the original coordinate system 
(i.e., as in the PDB file). Here, the indices from 0 to 3 are so 
labeled in the lexicographical order of their atom types.
When calculating the local coordinates of an atom in the refset, 
the origin is set to $\mathbf{r}_0$.
The $x$-axis is defined by the unit vector parallel to 
$\mathbf{r}_{01} \equiv \mathbf{r}_1 - \mathbf{r}_0$, 
that is, $\hat{\mathbf{x}} \equiv (1/\|\mathbf{r}_{01}\|)\mathbf{r}_{01}$.
With $\mathbf{r}_{02} \equiv \mathbf{r}_{2} - \mathbf{r}_0$, 
the $y$-axis is defined by $\hat{\mathbf{y}} \equiv (1/\|\mathbf{r}_{02}\|) \hat{\mathbf{x}}\times \mathbf{r}_{02}$. The $z$-axis is defined by 
$\hat{\mathbf{z}} \equiv \hat{\mathbf{x}} \times \hat{\mathbf{y}}$.
Thus, for a given set of coordinates $\mathbf{s}$ 
in the original system, the local coordinates in the system spanned by the 
refsets 
$\{\mathbf{r}_{i}\}$ are given as $\mathbf{s}' = 
[(\mathbf{s}-\mathbf{r}_0)\cdot\hat{\mathbf{x}}, 
(\mathbf{s}-\mathbf{r}_0)\cdot\hat{\mathbf{y}},
(\mathbf{s}-\mathbf{r}_0)\cdot\hat{\mathbf{z}}]$.
This coordinate system spanned by a refset is illustrated in 
Figure \ref{fig:refset}.
Using these notations, the definition of the chirality of a tetrahedron 
mentioned above is given as the sign of the dot product
$\mathbf{r}_{03}\cdot\hat{\mathbf{y}}$. For example, 
the chirality of the tetrahedron in Figure \ref{fig:refset} is positive.

\subsection{Extracting ligand binding sites}
\label{sec:lbsml}
By using the annotations in PDBML files, we identified the so-called hetero 
atoms (ligand atoms), and all protein atoms that are in contact with any of 
the hetero atoms. Two atoms are defined to be in contact if their distance
is less than or equal to 5\AA{}. 
For each ligand, we create an XML file 
containing a list of protein atoms that are in contact with it.
We call this XML file an LBSML file 
Atomic coordinates in an LBSML file are stored in the ``extatom'' style of 
the PDBML file\cite{PDBML} so that the ligand binding site can be examined 
visually by using the PDBjViewer \cite{KinoshitaANDNakamura2004}.
A set of protein atoms in contact with a ligand is called a ligand binding site.
We also calculate refsets of the PDB entry.
Along with the atomic coordinates of the ligand and the ligand binding site, 
the information of refsets and its type, volume, and lengths of edges of 
the tetrahedra defining the refsets is stored in an LBSML file. 
Refsets are saved in an LBSML file only if at least one of its vertex 
atoms is in contact with the ligand.
The distance threshold for the contact between refset and ligand atoms
was set to 5 \AA{}. 
As a result, we constructed 162,626 LBSML files corresponding to 
the ligand binding sites. 
A set of atoms in a ligand binding site is also referred to as a template 
in the following.

\subsection{Compilation of atomic coordinates and reference sets}
\label{sec:atomcord}
We compile the information of LBSML files
into tables of a relational database management system (RDBMS).
The use of RDBMS allows us to handle a huge amount of structural data 
relatively efficiently.
Basic information of LBSML files is saved in a table shown in 
Table \ref{tbl:lbsmldb}.
\begin{table}[tb]
\caption{\label{tbl:lbsmldb}Definition of the table for ligand binding sites.}
\begin{center}
\begin{tabular}[h]{l}\hline
\verb|CREATE TABLE lbsmldb (|\\
\verb|  lbsml_id INTEGER PRIMARY KEY, --- (a)|\\
\verb|  lbsml    TEXT,                --- (b)|\\
\verb|  pdbx     TEXT,                --- (c)|\\
\verb|  ligand   TEXT,                --- (d)|\\
\verb|  natoms   INTEGER );           --- (e)|\\\hline
\end{tabular}
\end{center}
(a) unique identifier; (b) file name; (c) PDB's description of the protein;
(d) PDB's annotation of the ligand; (e) the number of protein atoms in contact 
with the ligand.
\end{table}

Refsets in each LBSML file were compiled in a table (Table \ref{tbl:refsetdb}) 
along with their features such as tetrahedron type, volume, and edge lengths 
as well as the reference to the LBSML file they are derived from, 
and their serial number (refset identifier) in the LBSML file 
(note there are usually multiple refsets in a single LBSML file). 
There were about 4.7 million refsets in total. 
The primary key of this table consists of a pair of the reference to 
LBSML file and the refset identifier.
The types and local coordinates of atoms under each refset in an LBSML 
file are compiled into the same row as the refset.
\begin{table}[tb]
\caption{\label{tbl:refsetdb} Definition of the refset table.}
\begin{center}
  \begin{tabular}[h]{l}\hline
\verb|CREATE TABLE refsetdb (|\\
\verb|  lbsml_id  INTEGER,             --- (a)|\\
\verb|  irs       INTEGER,             --- (b)|\\
\verb|  PRIMARY KEY (lbsml_id,irs)     --- (c)|\\
\verb|  tetra     TEXT,                --- (d)|\\
\verb|  tvol      DOUBLE PRECISION,    --- (e)|\\
\verb|  td01      DOUBLE PRECISION,    --- (f)|\\ 
\verb|  td02      DOUBLE PRECISION,    --- (f)|\\ 
\verb|  td03      DOUBLE PRECISION,    --- (f)|\\ 
\verb|  td12      DOUBLE PRECISION,    --- (f)|\\
\verb|  td23      DOUBLE PRECISION,    --- (f)|\\
\verb|  td31      DOUBLE PRECISION,    --- (f)|\\
\verb|  atype_id  INTEGER [],          --- (g)|\\
\verb|  xco       DOUBLE PRECISION [], --- (h)|\\
\verb|  yco       DOUBLE PRECISION [], --- (h)|\\
\verb|  zco       DOUBLE PRECISION []  --- (h)|\\
\verb|);|\\
\hline
  \end{tabular}
\end{center}
(a) reference to ``lbsmldb'' (Table \ref{tbl:lbsmldb}); 
(b) reference set identifier;
(c) a pair of \verb|lbsml_id| and \verb|irs| makes the primary key of the refset.
(d) tetrahedron type; 
(e) volume of tetrahedron; 
(f) ``\verb|td|$ij$'' denotes the length of edge between vertices $i$ and $j$ of tetrahedron
(A tetrahedron consists of four atoms denoted $i, j=$ 0, 1, 2, and 3).
(g) types of the atoms spanned by the refset (encoded as integers).
(h) local coordinates of the atoms spanned by the refset.
\end{table}

For any database systems, it is critical to create appropriate indexes 
for efficient information retrieval. According to Garcia-Molina et al.\cite{DB_complete}, 
``an index is any data structure that takes as input a property of records --
typically the value of one or more fields -- and finds the records with 
that property `quickly.' ''
Here, we used an index based on the data structure called a 
B+ tree\cite{DB_complete}.
The refset table (Table \ref{tbl:refsetdb}) is indexed by the tetrahedron type, volume, and 
edge lengths with the SQL expression
``\verb|CREATE| \verb|INDEX| \verb|tetraIdx| \verb|ON| \verb|refsetdb| 
\verb|(tetra,| \verb|tvol,| \verb|td01,| \verb|td02,| \verb|td03,| 
\verb|td12,| \verb|td23,| \verb|td31)|.''

\subsection{The geometric indexing search method}
\label{sec:search}
Given a query protein structure, we search for ligand binding 
sites stored in the database that match a sub-structure of the query.
To do so, we first define and select the refsets (tetrahedra) 
of the query structure by the same procedure as the templates except that
contacts with hetero atoms are not taken into account (because they may not 
be present in the query structure). 
Then, for each refset of the query, we calculate the atomic 
coordinates of each atom under that 
refset. Next, we retrieve from the database those refsets 
whose the tetrahedron types are the same as that of the query tetrahedron, 
and whose volume and edge lengths are close to the corresponding 
quantities of the tetrahedron of the query within predefined threshold.
At the same time, those atomic coordinates which 
are based on the matching refsets are extracted from the database. 
This can be carried out with the SQL expression in Table \ref{tbl:fastsql}.
The retrieval of refsets and atomic coordinates are performed 
efficiently owing to the index constructed above. 
\begin{table}[tb]
  \caption{Pseudo SQL expression for local structure search.}
  \label{tbl:fastsql}
\begin{center}
  \begin{tabular}[h]{l}\hline
\verb|SELECT atype,xco,yco,zco,lbsml_id,irs|\\
\verb|FROM refsetdb|\\
\verb|WHERE AND tetra = '|$t_q$\verb|'| \\
\verb|  AND tvol BETWEEN | $v_q-\Delta_v$ \verb|AND| $v_q+\Delta_v$\\
\verb|  AND td01 BETWEEN |$d_{01} - \Delta_d$ \verb|AND| $d_{01} + \Delta_d$\\
\verb|  AND td02 BETWEEN |$d_{02} - \Delta_d$ \verb|AND| $d_{02} + \Delta_d$\\
\verb|  AND td03 BETWEEN |$d_{03} - \Delta_d$ \verb|AND| $d_{03} + \Delta_d$\\
\verb|  AND td12 BETWEEN |$d_{12} - \Delta_d$ \verb|AND| $d_{12} + \Delta_d$\\
\verb|  AND td23 BETWEEN |$d_{23} - \Delta_d$ \verb|AND| $d_{23} + \Delta_d$\\
\verb|  AND td31 BETWEEN |$d_{31} - \Delta_d$ \verb|AND| $d_{31} + \Delta_d$\\
\hline
  \end{tabular}
\end{center}
The table \verb|refsetdb| is defined in Table \ref{tbl:refsetdb}. 
$t_q$, $v_q$, and $d_{ij}$ are the type, volume, and edge length of a 
refset of the query. $\Delta$'s are predefined constants for similarity 
thresholds.
Expressions such as ``$v_q - \Delta_v$'' are given as constants in the actual code. We set $\Delta_v = 1$\AA{}$^{3}$ and $\Delta_d = 2$ \AA{}.
\end{table}
At this point, we have a list of tuples of atom type, coordinates, and 
LBSML file (\verb|lbsml_id|) and refset identifiers (\verb|refset_id|) 
returned by the SQL expression in Table \ref{tbl:fastsql}. 
Then, for each local atomic coordinates of the query, we select 
from the tuple list those tuples whose atom type is the same as that of 
the query and coordinates close to those of the query.
The query and template coordinates $(x_q, y_q, z_q)$ and $(x_t, y_t, z_t)$
are defined to be close if the distance between them is lower than 
a predefined constant $\Delta_c$ (Here we set $\Delta_c = 2$\AA{}).
Finally, the LBSML file and refset identifiers, on which the 
retrieved atomic coordinates are based, are recorded, 
and the count of the triple (template LBSML file, and query and template 
refset identifiers) is incremented.

After all the query refsets are examined, we have a list of tuples of 
a LBSML file, a template refset identifier and a query refset identifier, 
as well as the count of each tuple. 
If the count is sufficiently large, the local structure 
in the LBSML file is likely to be present in the query structure.
However, the count can be large just because there are a large number 
of atoms in certain templates. Therefore we use the score $S(f,r_t, r_q)$ 
of the tuple of LBSML file $f$, template refset identifier $r_t$ and 
query refset identifier $r_q$ defined as 
\begin{equation}
  \label{eq:score1}
  S_{GI}(f,r_t,r_q) = \frac{[\mathit{cnt}(f,r_t,r_q)]^p}{N_{f}}
\end{equation}
where $\mathit{cnt}(f,r_t,r_q)$ is the count of the tuple $(f,r_t,r_q)$
and $N_f$ is the number of atoms in the template of the LBSML file $f$.
We found that the best performance is attained with $p = 2$, and this value
is used throughout.
We refer to this score as the ``GI score'' (after Geometric Indexing) 
in the following.
The pairs of $(f,r_t,r_q)$ are sorted in the decreasing order of 
$S_{GI}(f,r_t,r_q)$, and the top $N_{top}$ hits (say, $N_{top} = 10000$) were 
saved for further refinement.

This search method, which we refer to as ``GI search'' in the following, 
is similar to the geometric hashing (GH) 
method\cite{GeometricHashing,NussinovANDWolfson1991}. 
However, it is not necessary to keep the database on memory, and 
atomic coordinates not not matched directly by using a hash function. 
Instead, we use a conventional RDBMS for keeping the template information, and 
first select matching template refsets using an index of the database.
In the present method, a matching refset serves not only as 
the basis of a local coordinate system but also as a seed alignment.

\subsection{Iterative refinement of alignment (IR procedure)}
\label{sec:refine}
By using the RDBMS-based search method, we can retrieve a set of 
ligand binding sites (and refsets) which are structurally 
similar to sub-structures of a query protein structure. 
At this point, however, the exact alignment of 
query and template atoms has not been obtained yet since all we have is the 
count of the tuple of LBSML files and template and query refset identifiers.
As in the GH method, it is possible to obtain an alignment by using a 
strict definition of the neighbor of an atom in the RDBMS-based method.
However, a small difference in the refsets could greatly perturb the quality 
of alignment. Therefore, it is desirable to 
employ a more robust method for refining the alignment at atomic resolution.

Since we assume that template and query atoms are approximately in 
the same refset, a reasonable set of possible alignments is obtained
by the following procedure. First we regard the system of query and template 
atoms as a bipartite graph\cite{Bollobas} in which query atoms form one 
group and template atoms another, and edges are allowed only between the 
two groups.
We assign an edge if the query atom $i$ and template atom $j$ are of the same atomic type and the distance $d_{ij}$ between them is less than 2 \AA{}. We 
assign a weight of $w_{ij} = 1 - d_{ij}/2$ to the edge. 
In an alignment, each query atom can match with at most 
one template atom.  The best alignment is the one for which 
the sum of the matching edges is larger than or equal to any other alignments.
This combinatorial optimization problem, called the maximum weight 
bipartite matching problem, can be readily solved by using the 
so-called Hungarian method\cite{Lawler,GuptaANDYing1999}.

The refinement of alignment is performed iteratively as follows.
First, by using the refset obtained by the RDBMS-based search, we 
construct a bipartite graph, and apply the Hungarian method to obtain the 
best matching (alignment). Second, we use the resulting alignment to 
rotate the template structure to optimally superpose onto the query structure.
This can be carried out by a classical least squares technique such as the 
quaternion-based one of Diamond\cite{Diamond1988}.
Third, based on the optimal superposition, we construct a new bipartite graph,
and apply the Hungarian method. The second and third stages are iterated 
until convergence which is achieved after 4 or 5 iterations on average.

The score of an alignment based on the LBSML file $f$, template refset identifier $r_t$ and query refset identifier $r_q$ is calculated as
\begin{equation}
  \label{eq:score2}
  S_{IR}(f,r_t,r_q) = \frac{ N_{ali}(f,r_t,r_q)\sum_{i,j}' w_{ij}}{N_f}.
\end{equation}
where the summation ($\sum'$) is over all the edges in the matching, 
$N_{ali}(f,r_t,r_q)$ is the number of aligned atom pairs and
$N_f$ is the number of atoms in the template of the LBSML file $f$.
We refer to this score as the ``IR score'' (after Iterative Refinement) 
in the following.

\subsection{Estimation of statistical significance}
In order to estimate the statistical significance of the IR score defined 
above, we introduce a statistical model based on random sampling.
After performing a GI search, we have a huge number of hits. Among those hits,
we randomly select 2,000 of them for iterative refinement.
As shown in the Results section, the distribution of the IR score of 
randomly selected alignments can be well approximated by a gamma distribution 
$\mathrm{GAM}(\alpha,\beta)$ whose probability density function is given as 
\begin{equation}
  \label{eq:gamma}
  f(x; \alpha,\beta) = \frac{1}{\beta\Gamma(\alpha)}
\left(\frac{x}{\beta}\right)^{\alpha-1}e^{-x/\beta}
\end{equation}
for $x \geq 0$ 
(note that the IR score is greater than or equal to 0 by definition).
Let the mean and variance of the IR scores of the randomly selected alignments
be $m$ and $v$, respectively. Then the parameters $\alpha$ and $\beta$ of the 
gamma distribution $\mathrm{GAM}(\alpha,\beta)$ are given as 
$\alpha = m^2/v$ and $\beta = v/m$, respectively.
Then the P-value or the probability that the IR score $T$ is greater 
than or equal to $x$ is given as 
\begin{equation}
  \label{eq:pval}
  P(S_{IR} \geq x) = \int_{x}^{\infty}f(x'; \alpha,\beta)\mathrm{d}x'
\end{equation}
which indicates that statistical significance of the IR score. That is,
lower P-values indicate greater statistical significance.
\subsection{Implementation}
\label{sec:imple}

All the codes were written in the Objective Caml (OCaml) language 
(http://caml.inria.fr). The RDBMS employed 
was the PostgreSQL system (http://www.postgresql.org) which has been 
moderately optimized for the underlying hardware.
All the computations were carried out on an Apple PowerMac 
(dual 2.5 GHz PowerPC G5) with 8 gigabytes (GB) RAM. 

\section{Results}
\label{sec:res}

\subsection{Execution time}
\label{sec:exectime}
We analyzed the execution time of a single search by using a mutant 
sperm whale myoglobin (PDB ID: 101m) as a query.
The number of hits subject to the refinement was set to 50,000.
The database consists of 162,626 ligand binding sites (LBSML files), 
4,699,804 refsets (tetrahedra).
In total, the hard disk space of 10 GB was consumed by the database.

The whole search process took 161 minutes of CPU time, in which 
115 minutes were spent for the GI search, 45 minutes for 
the IR procedure. In the GI search, the SQL expressions for selecting compatible
template refsets (Table \ref{tbl:fastsql}) took 90 minutes, and 
other parts took 25 minutes.
Thus, the execution of the SQL expression is the most time-consuming part of 
the whole process. This is because it involves access to the hard disk.
In the PDB entry 101m, there were 376 refsets selected according to the 
criteria described above. 
The search time is roughly proportional to the number of refsets of the query.
For each refset, an SQL expression for selecting compatible template refsets 
(see Table \ref{tbl:fastsql}) was issued.

\subsection{Effects of refinement}
The scores used in the geometric indexing and iterative refinement 
stages are different (see Eqs. \ref{eq:score1} and \ref{eq:score2}).
Accordingly, the rank of high-scoring templates may change 
between before and after the refinement.
To examine the effect of the refinement, we performed a search using 
the myoglobin (PDB ID: 101m) again. The top 50,000 
hits of the GI search were used for the refinement.

\begin{figure}[tb]
  \centering
  \includegraphics[width=8cm]{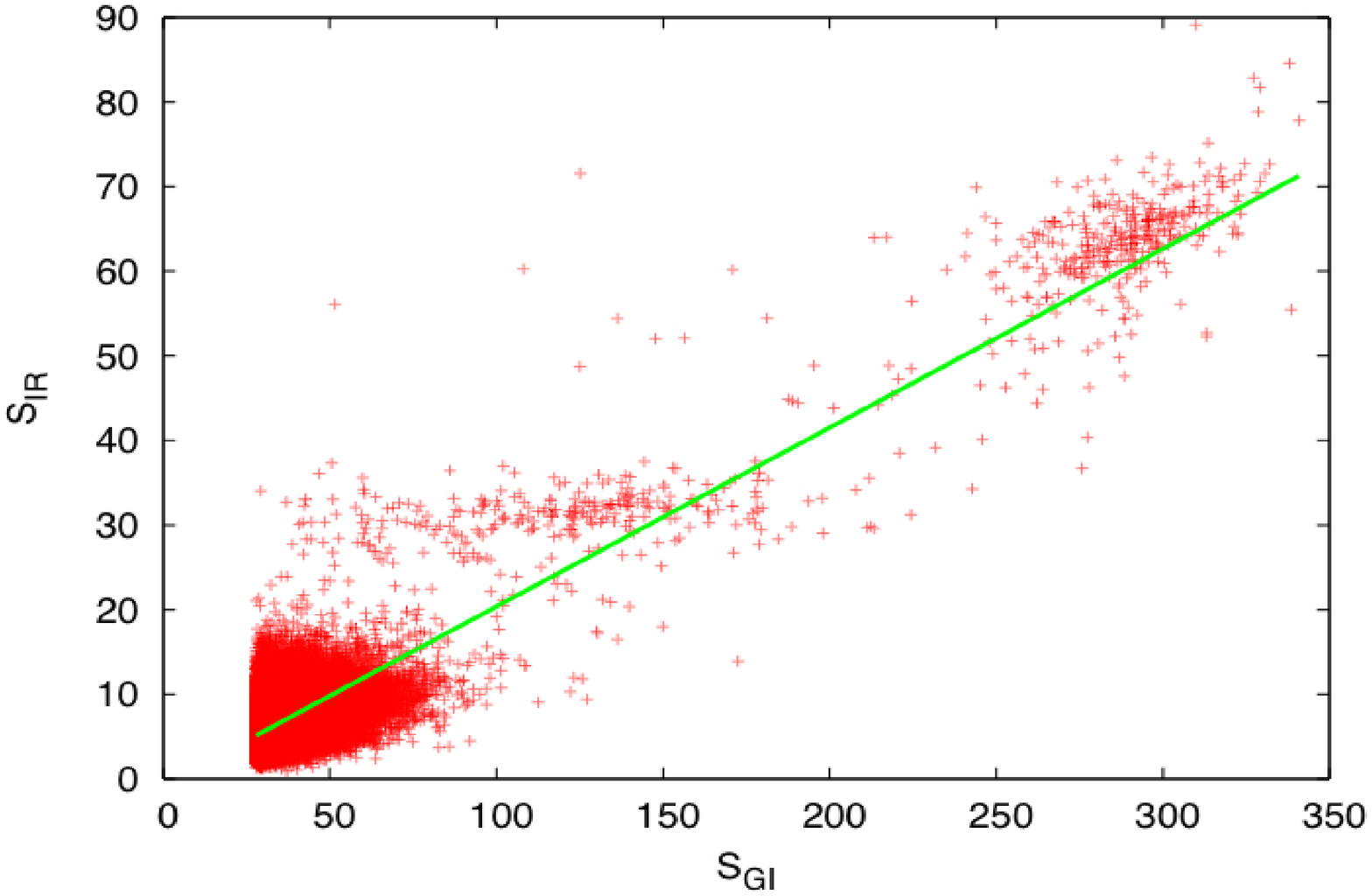}
  \caption{Comparison of GI score and IR score. Each point represents a 
template included in the top 50,000 hits for the query (PDB ID: 101m).
The regression line is also shown. The correlation coefficient 
between the scores is 0.87.}
  \label{fig:res50k}
\end{figure}

Figure \ref{fig:res50k} shows the two scores of each of the 50,000 templates. 
In general, the two scores correlate with each other very well, 
with a correlation coefficient of 0.87 in this case. 
But the rank of some templates may change dramatically upon refinement. 
The refinement greatly improved the scores of some templates of relatively 
low GI scores.

\subsection{Modeling the distribution of IR scores}
\begin{figure}[tb]
  \centering
  \includegraphics[width=8cm]{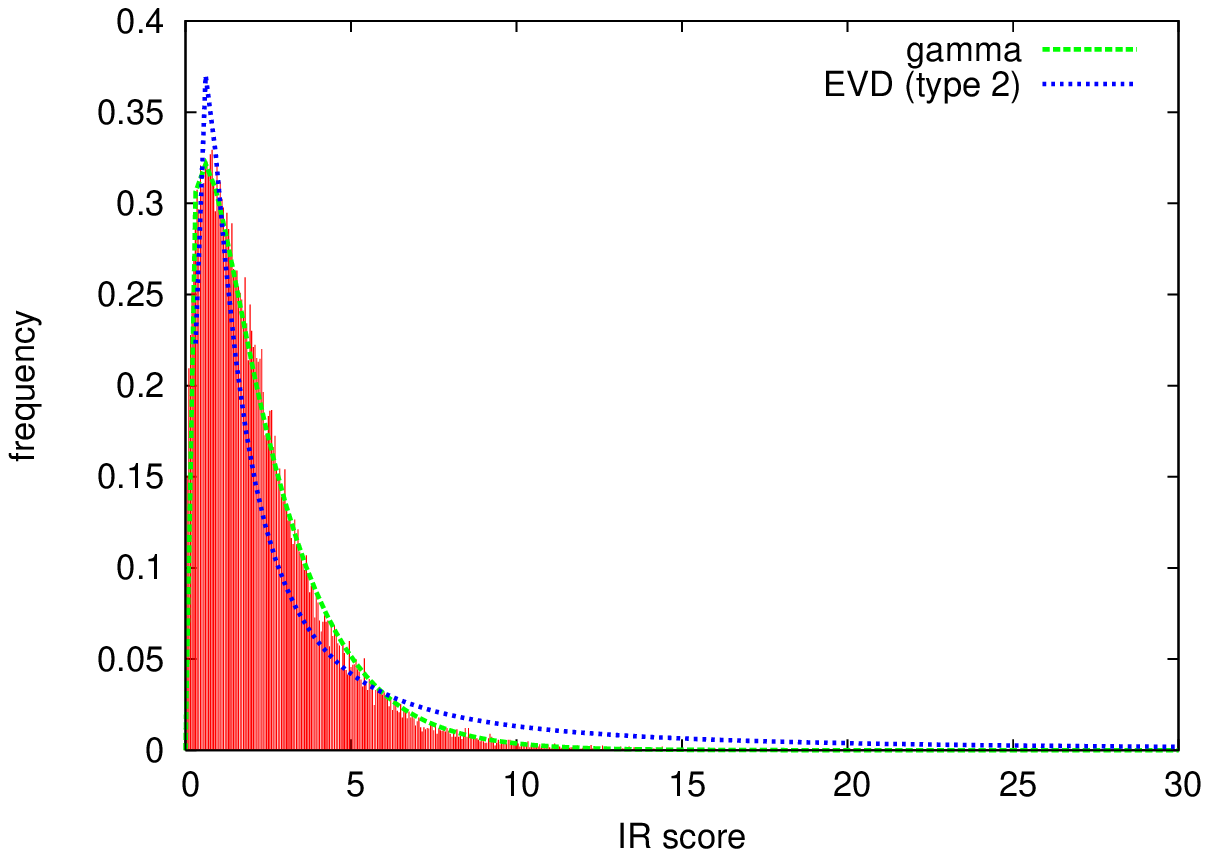}
  \caption{Distribution of IR scores of randomly selected templates.
The red bars indicate the histogram of IR scores of randomly selected templates 
obtained for the query 101m. The green line is the probability density 
function (PDF) of the gamma distribution $\mathrm{GAM}(\alpha,\beta)$ with 
the parameters $\alpha=1.32$ and $\beta=1.75$ calculated from 
the mean and variance of the scores. The blue line is the PDF of the type 2 
extreme value distribution with the parameters determined to best fit the 
histogram.}
  \label{fig:random}
\end{figure}

In order to estimate the statistical significance of IR score, 
we examined its distribution.
We first performed a GI search, and then randomly selected 50,000 
hits for iterative refinement. 
After the refinement, the histogram of the IR score was plotted.
Fig. \ref{fig:random} is an example obtained for the query 101m.
It is clearly seen that the distribution is well approximated 
by a gamma distribution (Fig. \ref{fig:random}, green line). 
We also fitted the type-2 (Fr\'echet) extreme value distribution 
(since the IR score is non-negative), but the fit was not as 
good as the gamma distribution (Fig. \ref{fig:random}, blue line).
The same trend was observed for other proteins.
Thus, we use the gamma distribution for calculating the statistical 
significance of the IR score.
Since the parameters of the gamma distribution may be different depending 
on queries, they are calculated by random sampling each time a search is 
performed.

\subsection{Examples of high-scoring alignments}
\label{sec:ex}

\paragraph{Myoglobin}
We first examine more closely the results obtained for the myoglobin (PDB ID:
101m) used above. We used the 50,000 hits by GI search for the 
further refinement.
The heme binding site of myoglobins occupied the first 363 hits with IR 
scores ($P$-values) ranging 
from 89.1 ($4.6\times 10^{-23}$) to 38.5 ($3.9\times 10^{-10}$).
Below the myoglobins were other globins such as hemoglobins 
and cytoglobins, all of which were identified by the heme binding sites.
The first non-globin appeared at the 555th rank with IR score of 30.1 
($P= 5.3\times 10^{-8}$). This entry was an isopropanol binding site of 
single-strand selective monofunctional uracil DNA glycosylase 
(UDG; PDB ID: 1oe6\cite{1OE6}). 
Visual inspection of the alignment suggests that this is likely to be
a false hit because the ligand binding site corresponds to inside an $\alpha$ 
helix. 
The next non-globin hit was the S-oxymethionine ``binding'' site of catalase 
(PDB ID: 2iuf\cite{2IUF}). S-oxymethionine here is actually a modified residue
in the protein which happened to be annotated as HETATM in the PDBML file.
This entry has a high score because the site is made of parts 
of $\alpha$ helices and $\alpha$ helices are common in globins. 
The next non-globin hit at the 489th rank with IR score of 26.1 ($P = 5.4\times 10^{-7}$) was 
a hypothetical protein from \emph{Pseudomonas aeruginosa} (PDB ID: 1tu9). 
Although its function is not well known, the fold of this protein is 
globin-like (Y. Kim et al., unpublished) and the aligned atoms 
comprised the heme-binding site.

\begin{figure}[tb]
  \centering
  \includegraphics[width=8cm]{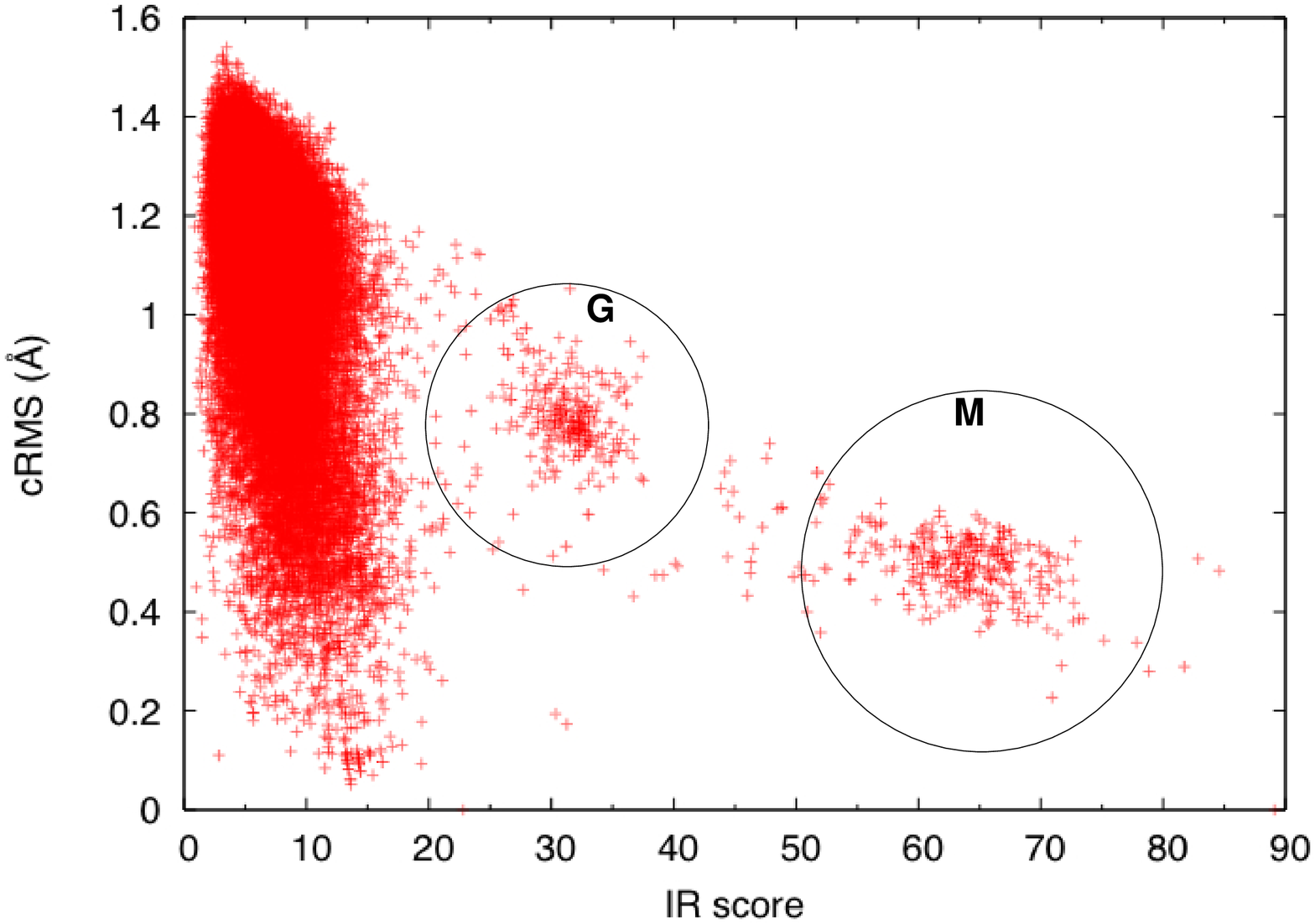}
  \caption{\label{fig:scat_101m}Scatter plot of the IR scores and coordinate 
RMS deviations resulted from a search with the PDB entry 101m. The regions 
enclosed by the circles marked with M and G contain mostly myoglobins and 
other globins, respectively.}
\end{figure}

In general, good alignments should have high IR scores and low coordinate 
root mean square (cRMS) deviations. This trend is clearly observed in 
Figure \ref{fig:scat_101m}. That is, good alignments should reside in 
the right bottom corner of the scatter plot of Figure \ref{fig:scat_101m}.
In this scatter plot, we can recognize two high-scoring clusters around 
IR score of 60--70 and 25-35,  which correspond to closely related myoglobins
and other globins, respectively.
In the region of low IR scores, there are may templates with low cRMS values.
A low IR score implies a small number of aligned atoms, hence the low cRMS 
values. 

\begin{figure}[tb]
  \begin{center}
    \includegraphics[width=14cm]{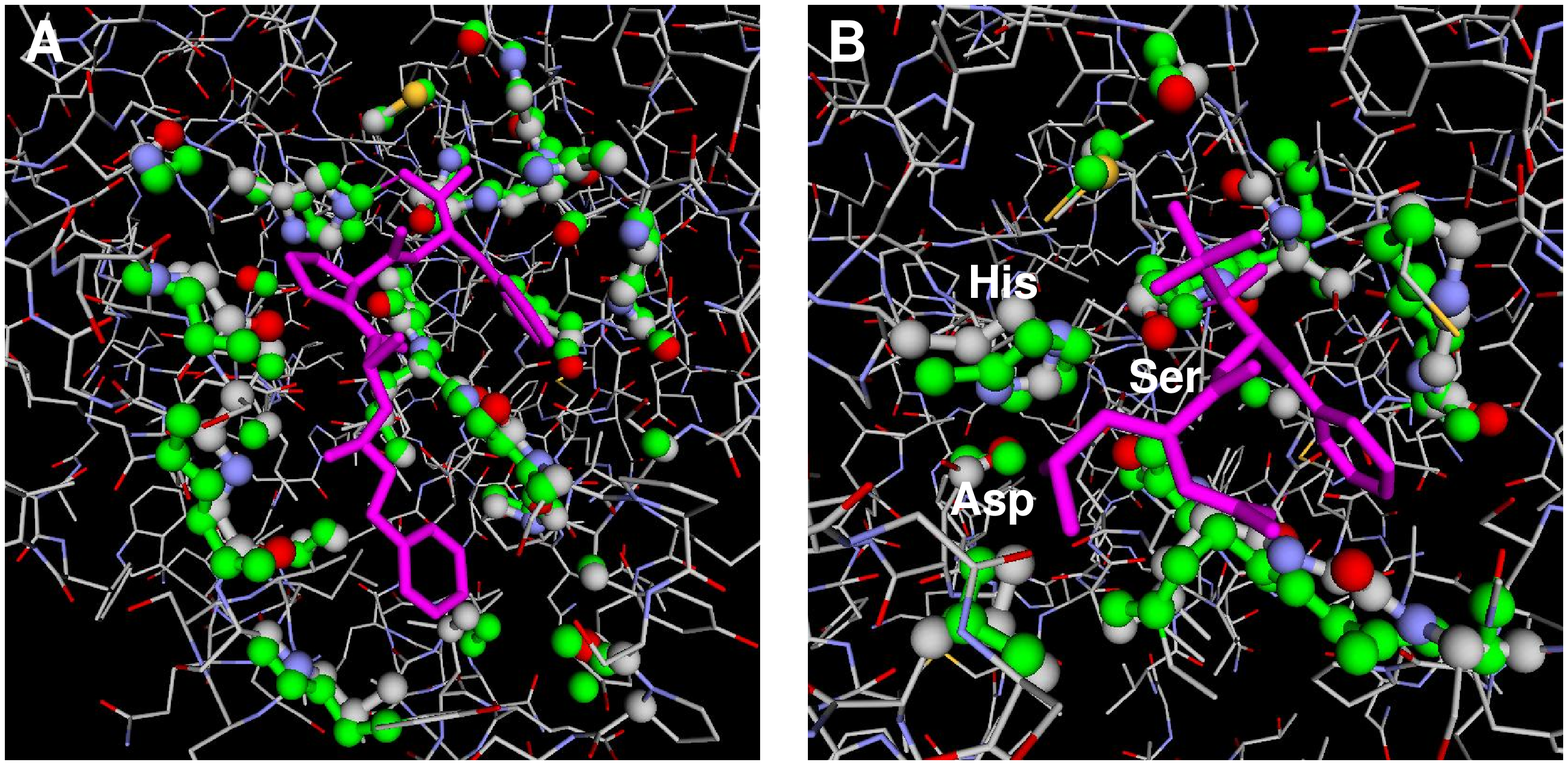}
  \end{center}
  \caption{Optimal superpositions of the query 1svn on templates. 
The wire-frame model in the CPK color scheme is the query protein 1svn. 
The template atoms are colored in green. Aligned atoms are in ball-and-stick
model. The ligand of the template is the ball-and-stick model in magenta.
A: Peptide-binding site of subtilisin DY (PDB ID: 1bh6\cite{1BH6}). 
B: Peptide-binding site of $\gamma$-chymotrypsin (PDB ID: 7gch\cite{7GCH}); 
the labeled Ser, His, Asp are the aligned catalytic triad. 
The figures were created by using 
the PDBjViewer\cite{KinoshitaANDNakamura2004}.}
  \label{fig:sup_1svn}
\end{figure}

\paragraph{Subtilisin savinase}
We next examine the result of a search with subtilisin savinase from \emph{Bacillus lentus} (PDB ID: 1svn\cite{1SVN}) as a query.
The top hit was the peptide binding site of subtilisin DY 
(PDB ID: 1bh6\cite{1BH6}) with an IR score of 59.8 and $P$-value of $1.0\times 10^{-14}$.
Subsequent hits were subtilisins and related proteases.
After these subtilisin-related templates (removing physically implausible 
templates), we found a Mn$^{2+}$ binding site of 
Dicer from \emph{Giardia intestinalis} (PDB ID: 2ffl\cite{2FFL}; 
$P = 1.5\times 10^{-5}$) and 
Mg$^{2+}$ binding site of 30S ribosomal 
subunit from \emph{Thermus thermophilus} (PDB ID: 1i94\cite{1I94}; 
$P = 1.8\times 10^{-5}$).
But these ion binding sites reside within common loop structures, and hence 
they are likely to be false positives.
At the 255th rank, we found the active site of bovine 
$\gamma$-chymotrypsin (PDB ID: 7gch\cite{7GCH}) with an IR score of 20.9 
($P$-value $2.0\times 10^{-5}$). 
This protein has a different fold than subtilisins but 
shares the common catalytic triad consisting of three residues 
Ser, His, and Asp. The obtained atomic alignment indeed contains these 
catalytic residues. Namely, Asp32, His64, and Ser221 of subtilisin Savinase are
aligned with Asp102, His57, and Ser195 of $\gamma$-trypsin 
(Fig. \ref{fig:sup_1svn} B).

\begin{figure}[tb]
  \centering
  \includegraphics[width=14cm]{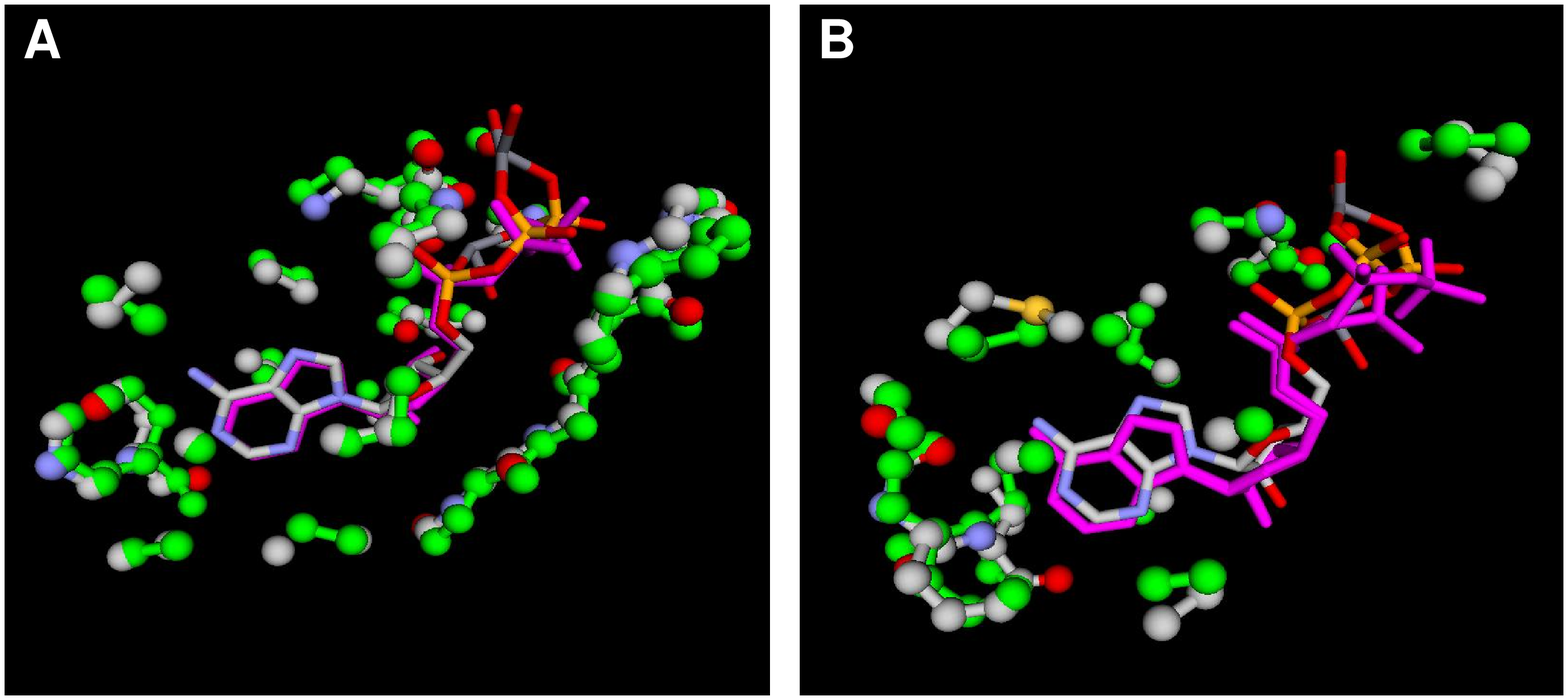}
  \caption{\label{fig:sup_1atp}
Optimal superpositions of the ATP-binding sites of the query cAMP-dependent 
protein kinase (cAPK; PDB ID: 1atp\cite{1ATP}) on templates. 
A: The template is the ATP-binding site of casein kinase-1 (PDB ID: 1csn\cite{1CSN}) from 
\emph{Schizosaccharomyces pombe}. 
B: The template is the ATP-binding site of glutathion synthetase (PDB ID: 1m0w\cite{1M0W}) from \emph{Saccharomyces cerevisiae}.
The color scheme is the same as 
Fig. \ref{fig:sup_1svn}. The ligand of 1atp is also shown in the stick model 
with the CPK colors.}
\end{figure}

\paragraph{cAMP-dependent protein kinase}
Our third example is the cAMP-dependent protein kinase, cAPK 
(PDB ID: 1atp\cite{1ATP}) from \emph{Mus musculus}. 
This example is motivated by the work of 
Kobayashi and Go\cite{KobayashiANDGo1997} where they have found that
the local structure of the nucleotide-binding site of cAPK is similar to 
those of other nucleotide-binding proteins with different folds.
They listed five ATP-binding proteins that share similar local structures:
glutaminyl-tRNA synthetase, D-Ala:D-Ala ligase (DD-ligase), 
casein kinase-1 (CK-1), 
seryl-tRNA synthetase, and glutamine synthetase\cite{KobayashiANDGo1997}. 
According to the SCOP database\cite{SCOP}, CK-1 and cAPK belong 
to the same family, the protein kinase catalytic subunit family, 
although the sequence identity between them is as low as 19\%.
Among the five proteins listed by Kobayashi and Go, 
CK-1 exhibited a highly significant similarity with an IR score of 42.8 and 
$P = 8.9\times 10^{-11}$ (Fig. \ref{fig:sup_1atp} A).
In contrast, we only found a weak similarity with glutathion synthetase, 
belonging to the same superfamily as DD-ligase, with a relatively low IR score 
of 12.5 ($P = 2.1\times 10^{-3}$).
Most high-scoring templates were all kinases of the same fold.
Other similarities listed by Kobayashi and Go were either not detected, 
or detected with wrong alignments. 
There are at least two possible explanations for this failure in detecting 
similar local structures. First, our criteria for selecting similar refsets 
may be too stringent so that possible hits are discarded during the GI search. 
Second, the number of aligned atoms as obtained by Kobayashi and Go is very 
small, ranging from 14 to 16, whereas some of obvious false hits contained 
more than 20 aligned atoms.
The first point may be corrected by loosening the criteria at the cost of 
increased CPU time. The second point is more problematic, however.
Kobayashi and Go used only ATP-binding proteins for their study while we used
all the ligand-binding sites present in the current PDB. Accordingly, 
the signal-to-noise ratio is substantially lower in the present case. 
In order to overcome this problem, a more elaborate statistical method 
may be necessary.

\begin{figure}[tb]
  \centering
  \includegraphics[width=14cm]{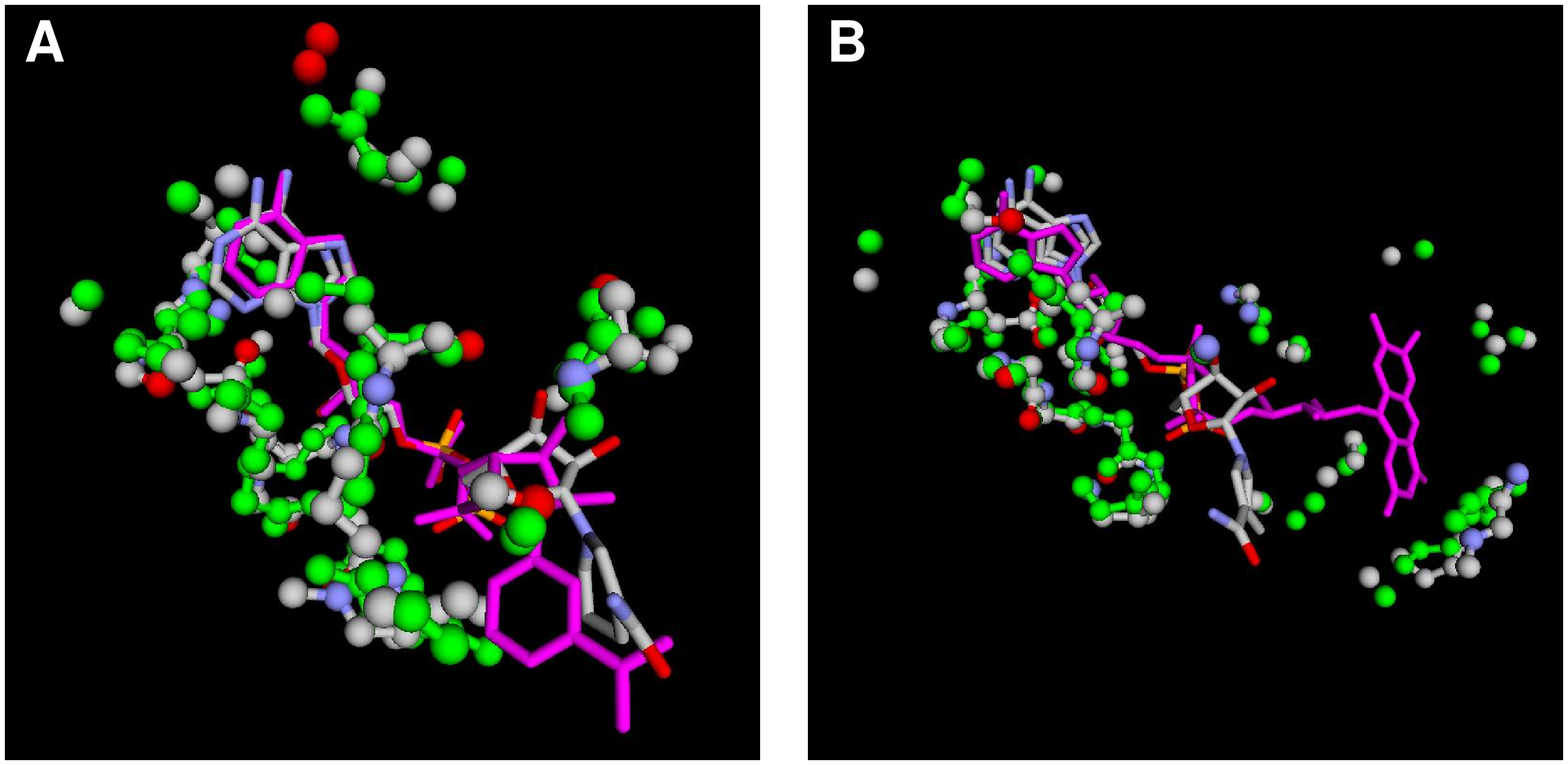}
  \caption{\label{fig:sup_1het}
Optimal superpositions of the NAD-binding 
sites of the query alcohol dehydrogenase (PDB ID: 1het)\cite{1HET} 
on templates. 
A: The template is the NAD-binding site of urocanase protein 
(PDB ID: 1x87; Tereshko et al., unpublished) from \emph{Bacillus stearothermophilus}. 
B: The template is the FAD-binding site of p-hydroxybenzoate hydroxylase (PDB ID: 1iuv\cite{1IUV}) from \emph{Pseudomonas aeruginosa}.
The color scheme is the same as Fig. \ref{fig:sup_1svn}. 
The ligand of 1het is also shown in the stick model with the CPK colors.}
\end{figure}

\paragraph{Alcohol  dehydrogenase}
The fourth example is the alcohol dehydrogenase (ADH; PDB ID: 1het\cite{1HET}) 
from \emph{Equus caballus} (horse).
The first 107 top hits are the nicotinamide-adenine-dinucleotide (NAD)-binding 
sites of ADHs from various species, which are followed by 
various kinds of other dehydrogenases such as formaldehyde dehydrogenase, 
sorbitol dehydrogenase, glucose dehydrogenase, and so on.
We looked for structural similarities with proteins other than dehydrogenases,
and have found a few such examples.
One example is  the NAD-binding site of the urocanase protein 
(PDB ID: 1x87; Tereshko et al., unpublished) with an IR score of 24.0 
($P= 2.7\times 10^{-6}$). According to the SCOP database, this protein belongs 
to the urocanase fold which is clearly different from the NAD(P)-binding 
Rossmann-fold domain of the ADH. The alignment (Fig. \ref{fig:sup_1het} A) 
consists of 76 atom pairs yielding cRMS of 1.0 \AA{}.
Another example is the flavin-adenine dinucleotide (FAD)-binding site of 
p-hydroxybenzoate hydroxylase (PHBH; PDB ID: 1iuv\cite{1IUV}) which 
exhibited a significant IR score of 20.2 ($P = 2.3\times 10^{-5}$).
PHBH belongs to the FAD/NAD(P)-binding domain fold which is 
different from the NAD(P)-binding Rossmann fold of ADH.

\section{Discussion}
We have demonstrated that the present method can detect non-trivial similarities
in protein local structures at atomic resolution in a reasonable CPU time.
Here we discuss a few remaining issues to be solved and possibilities for 
further improvements.

\subsection{Recurring false positives}
It was often observed that certain ligand binding sites exhibited high scores 
regardless of query structures. 
Such examples include the isopropanol binding site of UDG and the 
S-oxymethionine binding site of catalase as mentioned above in the example 
of myoglobin. 
These and other recurring false hits are almost always part of super-secondary 
structures which consist of $\alpha$-helices and $\beta$-strands which are
highly regular and abundant. 
Another source of error is the ambiguous definition of ``ligands.'' 
For example, the ligand in the S-oxymethionine binding site of 
catalase (2iuf\cite{2IUF}) described above is actually a modified residue
in the protein, not another molecule than the protein itself.
In this case, most part of the ligand (S-oxymethionine) should be treated 
as a part of the protein. Many of the ligands treated in this study 
are biologically irrelevant but are present as a part of the solvent.
Such examples include the isopropanol in the PDB entry 1oe6\cite{1OE6} described 
above.
Therefore, it would be helpful to include only biologically relevant ligands 
in the database although this may require a great deal of effort in the absence
of proper annotations.

\subsection{Increasing sensitivity}
In the proposed method, we first select candidates based on the attributes of 
refsets, such as the volume and edge length of tetrahedra. 
In the current implementation, the criteria for refsets are 
relatively stringent so that it is not guaranteed that all the possibly 
important refsets are stored in the database 
(e.g., tetrahedra containing multiple atoms of the same type). 
This may be a reason why the present method failed to detect some of the 
known similarities between cAPK and other proteins of different folds.
In order not to miss such important refsets, 
it may be possible to use backbone-based refsets\cite{PennecANDAyache1998}. 
However, the naive definition of backbone-based refsets (defined by three atoms 
N, C$_\alpha$, C) is extremely inefficient because all such refsets are 
essentially identical and we have to retrieve all such refsets every time we 
issue an SQL query similar to that of Table \ref{tbl:fastsql}.
Therefore, we need to add some extra attributes to efficiently select 
relevant candidates for retaining efficiency. For example, we may use 
similarity between amino acid residues or backbone dihedral angles for 
restricting possible candidates.

A better statistical model may also improve the sensitivity. 
Currently we employ a simple gamma distribution that depend only on the 
IR score. However, we observed that the IR score depends on cRMS in a 
systematic manner so that some false hits with relatively high IR scores with 
large cRMS values may be eliminated. 
Therefore, it may be helpful to estimate the 
cRMS-dependent parameters for the gamma distribution.

\subsection{Improving efficiency}
The method presented here can be relatively efficiently executed on a small 
desktop computer.
The key idea is to use a conventional RDBMS to handle the large amount of 
structural data. The most time-consuming part is the access 
to data stored on a hard disk.
Conventional RDBMS implements a cache mechanism so that frequently accessed 
data are stored in memory when possible. Using this mechanism, it is possible
speed up the similarity search by simply implementing the GI method in 
a computer with a large memory. This will automatically lead to the 
efficiency comparable to the GH method. 
However, unlike naive implementations of the GH method, 
the present GI method does not break even when the data size grows to such 
an extent that it does not fit into the memory. 

Another possible improvement may be made by reducing the number of 
query refsets to be examined. The current implementation requires a CPU 
time proportional to the number of refsets of the query, which 
ranges from $\sim$100 to 2,000 or more in typical proteins.
In the examples given above, a search with myoglobin (PDB ID: 101m) 
with 376 refsets took 
approximately 160 minutes while a search with alcohol dehydrogenase
(PDB ID: 1het) with 1654 refsets took 730 minutes ($\sim$12 hours).
If we can eliminate many of the query refsets which are unlikely to be
ligand binding sites, the computational time may be greatly reduced.

\section{Conclusion}
\label{sec:conclusion}
We have developed a method for searching for local atomic structures of 
proteins in database that are structurally similar to sub-structures of 
a given query protein structure.
In particular, we presented techniques based on a conventional 
relational database management system to practically deal with the 
huge amount of structural data currently available in the Protein Data Bank.
In spite of the facts that the size of the database is massive and that 
the resolution of the alignments obtained by the method is of the atomic level, 
the present method can yield search results typically within a few hours 
using an ordinary desktop computer. With further improvements discussed above,
the present method seems to be a 
promising approach to routinely searching for local structural similarity 
at atomic resolution, 
and to functional annotation of newly determined protein structures.
Finally it is noted that the core idea of the present method is a very 
general one, and is obviously applicable to other similar problems such as,
for example, the similarity search of molecular 
surfaces\cite{KinoshitaANDNakamura2003,Shulman-PelegETAL2004} where the 
geometric hashing 
technique is applicable in principle, but prohibitive in practice due 
to a huge data size.

\section*{Acknowledgments}
The authors thank Drs. Daron Standley and Kengo Kinoshita 
for critical comments on an early version of the manuscript.
This work was supported by a grant-in-aid from the Institute for Bioinformatics 
Research and Developmenet, the Japan Science and Technology Agency.

\bibliographystyle{biophysics}
\bibliography{refs,mypaper}
\end{document}